\documentclass[fleqn]{2026SCGE}
\usepackage{booktabs}
\usepackage{amsmath}
\usepackage{bm}
\usepackage{xcolor}
\newcommand{\Eq}[1]{equation~\eqref{#1}}
\newcommand{\dif}{\mathrm{d}}

\newcommand{\GW}{\mathrm{GW}}
\newcommand{\Pzeta}{\mathcal{P}_{\zeta}}
\newcommand{\fpbh}{f_{\mathrm{PBH}}}
\newcommand{\Cc}{\mathcal{C}_{\mathrm{c}}}
\newcommand{\Cg}{\mathcal{C}_{\mathrm{g}}}

\begin{document}
\ensubject{subject}

\ArticleType{Article}
\SpecialTopic{SPECIAL TOPIC: }
\Year{2026}
\No{1}
\DOI{??}
\ArtNo{000000}

\title{Projected Constraints on Primordial Black Holes from Scalar-Induced Gravitational Waves with Taiji}{Projected Constraints on Primordial Black Holes from Scalar-Induced Gravitational Waves with Taiji}

\author[1,2]{Yang Jiang}{jiangy@ucas.ac.cn}
\author[3]{Chen Yuan}{yuanchen@shu.edu.cn}
\author[1,2,4]{Qing-Guo Huang}{huangqg@itp.ac.cn}


\address[1]{School of Fundamental Physics and Mathematical Sciences,\\ 
Hangzhou Institute for Advanced Study, University of Chinese Academy of Sciences (UCAS), Hangzhou 310024, China}
\address[2]{School of Physical Sciences,\\
  University of Chinese Academy of Sciences,
  No. 19A Yuquan Road, Beijing 100049, China}
\address[3]{Department of Physics, Shanghai University, Shanghai 200444, China}
\address[4]{Institute of Theoretical Physics,\\
  Chinese Academy of Sciences, Beijing 100190, China}

\abstract{Scalar-induced gravitational waves (SIGWs) provide a direct probe of the enhanced primordial curvature perturbations that may also produce primordial black holes (PBHs). We forecast the capability of the space-based gravitational-wave observatory Taiji to search for an SIGW background generated by a broken-power-law curvature spectrum. A signal-injection study is used to validate the analysis pipeline, after which a pure-noise realization is employed to derive projected upper limits on the curvature-spectrum parameters. We translate these limits into constraints on the PBH abundance using the nonlinear compaction function, critical collapse, and the joint Gaussian distribution of the compaction amplitude and curvature at its peak. The resulting projected $95\%$ upper limits on the PBH dark-matter fraction satisfy $\fpbh^{95\%}<1$ over PBH masses from approximately $6.2\times10^{-18}\,M_\odot$ to $1.7\times10^{-8}\,M_\odot$. We compare the forecast with representative Hawking-evaporation and microlensing bounds. In part of the asteroid-mass interval, the projected Taiji limit is more restrictive than the current Subaru Hyper Suprime-Cam (HSC) microlensing constraint.}

\keywords{Primordial black hole, Scalar-induced gravitational wave, Taiji}

\maketitle
\begin{multicols}{2}
\section{Introduction}
The nature of dark matter remains one of the central problems in cosmology and particle physics. Despite extensive experimental searches~\cite{BOZORGNIA2025671,Xia2026ProgressAP,Baudis:2025yva}, no compelling non-gravitational signal of a particle dark-matter candidate has been established. Primordial black holes (PBHs), which can form from the collapse of sufficiently large density perturbations in the early Universe, provide an alternative candidate~\cite{1975Natur.253..251C,10.1093/mnras/152.1.75,Green:2024bam}. Their existence would also affect reionization and the formation of cosmic structure~\cite{Djorgovski:2004hc,Jhurani:2023tol,Yin:2026hvw,Garcia-Bellido:2017fdg,Atrio-Barandela:2022jov,Kovetz2025}. Complementary GW probes of PBHs include compact binary mergers and the GW signatures during the formation of PBHs~\cite{Chen:2021nxo,Huang:2024wse,Yuan:2024yyo,Yuan:2025avq,Yuan:2025bdp}.

Gravitational-wave (GW) observations offer a particularly direct route to the small-scale primordial fluctuations relevant for PBH formation. During radiation dominated era, scalar perturbations source tensor modes at second order through nonlinear mode coupling~\cite{10.1143/PTP.37.831,PhysRevD.47.1311,Ananda2006CosmologicalGW,Domenech:2021ztg,PhysRevD.107.063510,Yuan:2025seu}. These scalar-induced gravitational waves (SIGWs) retain information about the amplitude and shape of the primordial curvature spectrum. Their theoretical description has been explored through many aspects, including the analyses of the infrared scaling, higher-order corrections, the gauge behavior and the non-Gaussian effects~\cite{Yuan:2019udt,Yuan:2019wwo,Yuan:2019fwv,Yuan:2024qfz,Yuan:2025seu,Yuan:2020iwf,Meng:2022ixx,Yuan:2023ofl}. During the formation of PBHs, the same enhanced fluctuations will inevitably generate SIGWs. 

Searches for SIGWs have already constrained the primordial spectrum and the associated PBH abundance over several frequency ranges~\cite{Chen:2019xse,Yuan:2021qgz,Kapadia_2021,Jiang_2024,LISACosmologyWorkingGroup:2025vdz,Bian:2025ifp,Liu:2023pau,Bi:2026zlt}. In parallel, microlensing, compact-binary searches, evaporation products, and other astrophysical and cosmological observations have placed strong, but model-dependent, upper limits on the PBH dark-matter fraction~\cite{Mroz:2024mse,Mroz:2024wag,LIGOScientific:2025vwc,PhysRevLett.132.151401,Yuan:2024yyo}. In the asteroid-mass window, the strongest Subaru/HSC and OGLE microlensing limits reach $\fpbh\sim\mathrm{few}\times10^{-3}$~\cite{Niikura:2017zjd,Mroz:2024mse,Mroz:2024wag}. Under Gaussian initial statistics, PTA constraints on SIGWs challenge subsolar mass PBH as main dark matter \cite{Chen:2019xse} while negative non-Gaussianity is required for subsolar mass PBH as dark matter \cite{Liu:2023ymk,Franciolini:2023pbf}. The asteroid-mass interval remains among the least constrained regions, and it is naturally connected to SIGWs in the millihertz band.

Space-based interferometers such as the Laser Interferometer Space Antenna (LISA)~\cite{2017arXiv170200786A} and Taiji~\cite{LUO2020102918} are designed to operate in the millihertz band. Their sensitivity to stochastic backgrounds makes them suitable for testing curvature perturbations on scales that are inaccessible to the cosmic microwave background and large-scale-structure observations. 
For instance, a concrete non-minimal spectator-field realization produces asteroid-mass PBHs together with SIGWs in the sensitivity bands of Taiji, TianQin, and LISA~\cite{Meng:2022low}.

In this work, we forecast the SIGW and PBH constraints with a Taiji-like configuration. We adopt a broken-power-law (BPL) parametrization of the curvature spectrum, which covers a wide range of models ~\cite{Li:2024lxx}. We calculate the corresponding SIGW spectrum, perform Bayesian signal-recovery and null-signal analyses with simulated Taiji data, and convert the upper limits on the curvature spectrum into bounds on the PBH abundance. For this last step, we use the compaction-function formalism with the joint Gaussian distribution of the peak amplitude and curvature.

\section{Scalar-induced gravitational waves and PBH abundance}
\label{sec:theory}

\subsection{Scalar-induced gravitational waves}
In many inflation models, a localized enhancement of the comoving curvature spectrum can be represented by the BPL form
\begin{equation}
  \Pzeta(k)=A\,
  \frac{\alpha+\beta}
  {\beta(k/k_*)^{-\alpha}+\alpha(k/k_*)^{\beta}},
  \label{eq:bpl}
\end{equation}
where $A=\Pzeta(k_*)$ is the peak amplitude, $k_*$ is the characteristic scale, and $\alpha,\beta>0$ are the infrared and ultraviolet spectral indices, respectively.
For the data analysis, we use the frequency $f_*=k_*/(2\pi)$, where we set $c=1$.

The BPL template is adopted here for physical rather than merely phenomenological reasons. In many single-field inflationary scenarios that produce PBHs, the inflaton transiently departs from slow roll, often through an ultra-slow-roll or more general non-attractor stage, and the resulting enhancement of $\Pzeta$ has an approximately power-law rise and fall on the two sides of a localized peak~\cite{Ballesteros:2017fsr,Byrnes:2018txb,Carrilho:2019oqg,Atal:2021jyo,Karam:2022nym}. The two slopes in \Eq{eq:bpl} therefore provide a compact description of the peak width and asymmetry and these features would leave characteristic imprints on the SIGW spectrum. Moreover, a near-peak approximation method is developed in ~\cite{Li:2024lxx} to obtain an analytical description of the SIGWs generated by a BPL power spectrum, allowing rapid and accurate SIGW spectrum generation in data analysis. The BPL family is thus a physically motivated description to represent a broad class of PBH-producing scenarios. However, we emphasize that the present analysis is a template-based search rather than a model-independent reconstruction~\cite{LISACosmologyWorkingGroup:2025vdz}. Consequently, the limits reported in the following are conditional on the BPL family and should not be interpreted as general bounds on an arbitrary curvature power spectrum.

For Gaussian scalar perturbations, the dimensionless tensor power spectrum generated during radiation domination is~\cite{Kohri:2018awv}
\begin{multline}
  \overline{\mathcal{P}_h(k)}=4\int_0^\infty \dif v
  \int_{|1-v|}^{1+v}\dif u\,
  \left[\frac{4v^2-(1+v^2-u^2)^2}{4uv}\right]^2 \\
  \overline{I^2(u,v)}\,
  \Pzeta(ku)\Pzeta(kv),
  \label{eq:ph}
\end{multline}
where the oscillation-averaged kernel for radiation dominated era is~\cite{Espinosa:2018eve,Kohri:2018awv}
\begin{multline}
 \overline{I^2(u,v)}=
 \frac{9(u^2+v^2-3)^2}{32u^6v^6}
 \Bigg\{
 \bigg[-4uv+(u^2+v^2-3)\cdot \\
 \ln\left|\frac{3-(u+v)^2}{3-(u-v)^2}\right|\bigg]^2
 +\pi^2(u^2+v^2-3)^2\Theta(u+v-\sqrt{3})
 \Bigg\}.
 \label{eq:kernel}
\end{multline}
The present-day fractional energy density per logarithmic wavenumber interval is
\begin{equation}
  \Omega_{\GW,0}(k)=
  \frac{\Omega_{r,0}}{24}
  \left(\frac{g_*}{g_*^0}\right)
  \left(\frac{g_{*s}^0}{g_{*s}}\right)^{4/3}
  \overline{\mathcal{P}_h(k)},
  \label{eq:omega_gw}
\end{equation}
where $\Omega_{r,0}=9\times10^{-5}$, and $g_*$ and $g_{*s}$ ($g_*^0$ and $g_{*s}^0$) denote the effective relativistic degrees of freedom for the energy and entropy densities at horizon re-entry (at the present epoch).

\subsection{PBH formation from the curvature spectrum}
\label{subsec:pbh}

We calculate the PBH abundance using threshold statistics applied to local maxima of the nonlinear compaction function. This construction combines the long-wavelength treatment of nonlinear cosmological perturbations with Gaussian peak statistics and subsequent refinements accounting for the nonlinear curvature--density relation, the profile dependence of the collapse threshold, and critical scaling near the threshold~\cite{Shibata:1999zs,Harada:2015yda,Bardeen:1985tr,Young:2019yug,DeLuca:2019qsy,Kawasaki:2019mbl,Ferrante:2022hwi,Ianniccari:2024jsf}.

For an approximately spherical perturbation, the nonlinear compaction function at leading order in the gradient expansion can be written as~\cite{Harada:2015yda,Young:2019yug,DeLuca:2019qsy,Kawasaki:2019mbl}
\begin{equation}
  \mathcal{C}(r)=\Cg(r)-\frac{1}{4\Phi}\Cg^2(r),
  \quad
  \Cg(r)=-2\Phi r\zeta'(r),
  \quad
  \Phi=\frac{3(1+w)}{5+3w},
  \label{eq:compaction}
\end{equation}
where $w$ is the equation-of-state parameter. During radiation domination, $w=1/3$ and $\Phi=2/3$, so that $\mathcal{C}=\Cg-3\Cg^2/8$. We denote by $r_m$ the radius at which the compaction function reaches a local maximum. For type-I perturbations, $\Cg(r_m)<2\Phi$, the maximum of $\mathcal{C}$ coincides with that of $\Cg$, and~\cite{Ferrante:2022hwi,Ianniccari:2024jsf}
\begin{equation}
  \mathcal{C}''(r_m)=\Cg''(r_m)
  \left[1-\frac{\Cg(r_m)}{2\Phi}\right].
  \label{eq:compaction_second}
\end{equation}

A PBH forms if $\mathcal{C}(r_m)$ exceeds the collapse threshold $\Cc$. The threshold is not a universal constant when expressed in terms of the compaction at its maximum, but depends on the shape and curvature of the perturbation profile~\cite{Harada:2013epa,Germani:2018jgr,Musco:2018rwt,Escriva:2019nsa,Musco:2020jjb}. For each BPL curvature spectrum, we construct the representative mean curvature profile, locate the maximum of its compaction function at $r_m$, and determine the corresponding spectrum-dependent collapse threshold $\Cc$ using the semi-analytic prescription of~\cite{Musco:2020jjb}. This prescription relates the shape of the primordial power spectrum to the shape parameter of the mean compaction profile and incorporates calibrations from numerical-relativity simulations. Thus, $\Cc$ is fixed by the assumed BPL spectral shape rather than treated as a universal constant or varied independently for each realization entering the abundance integral.

For Gaussian curvature perturbations, the variables
\begin{equation}
  X\equiv-\frac14 r_m^2\Cg''(r_m),
  \quad
  Y\equiv\Cg(r_m),
  \label{eq:XY}
\end{equation}
are linear functionals of the Gaussian field $\zeta$ and therefore follow a joint Gaussian distribution. Introducing $\bm{v}=(X,Y)^{\mathrm{T}}$, their probability density is~\cite{Bardeen:1985tr,Yoo:2020dkz,Ferrante:2022hwi,Ianniccari:2024jsf}
\begin{equation}
  P(X,Y)=\frac{1}{2\pi\sqrt{\det\bm{\Sigma}}}
  \exp\left(-\frac12\bm{v}^{\mathrm{T}}
  \bm{\Sigma}^{-1}\bm{v}\right),
  \qquad
  \bm{\Sigma}=\begin{pmatrix}
      \sigma_2^2 & \sigma_1^2\\
      \sigma_1^2 & \sigma_0^2
  \end{pmatrix},
  \label{eq:joint_pdf}
\end{equation}
with
\begin{equation}
  \sigma_0^2=\langle \Cg^2\rangle,
  \quad
  \sigma_1^2=-\frac14 r_m^2\langle\Cg''\Cg\rangle,
  \quad
  \sigma_2^2=\frac1{16}r_m^4\langle(\Cg'')^2\rangle.
  \label{eq:covariances}
\end{equation}

The covariance matrix is determined directly by the primordial spectrum. With a real-space top-hat window, whose Fourier-space form is
\begin{equation}
  W(x)=3\frac{\sin x-x\cos x}{x^3},
  \qquad x=kr_m,
\end{equation}
the radiation-era linear compaction in Fourier space is
\begin{equation}
  \Cg(\bm{k},r_m)=K_0(x)\zeta(\bm{k}),
  \qquad
  K_0(x)=\frac49 x^2W(x).
  \label{eq:Cg_fourier}
\end{equation}
The choice of smoothing window is part of the PBH-abundance prescription and constitutes a systematic uncertainty in the mapping from $\Pzeta$ to $\fpbh$~\cite{Ando:2018qdb,Young:2019gfc,Yoo:2020dkz}. Defining
\begin{equation}
  K_2(x)=-\frac{x^2}{4}\frac{\dif^2K_0}{\dif x^2},
\end{equation}
the covariance elements can equivalently be evaluated as
\begin{equation}
  \begin{aligned}
  \sigma_0^2&=\int_0^\infty\frac{\dif k}{k}\,
  K_0^2(kr_m)\Pzeta(k),\\
  \sigma_1^2&=\int_0^\infty\frac{\dif k}{k}\,
  K_0(kr_m)K_2(kr_m)\Pzeta(k),\\
  \sigma_2^2&=\int_0^\infty\frac{\dif k}{k}\,
  K_2^2(kr_m)\Pzeta(k).
  \end{aligned}
  \label{eq:spectral_moments}
\end{equation}
The probability density $P[\mathcal{C}(r_m),\mathcal{C}''(r_m)]$ is obtained from \Eq{eq:joint_pdf} through the nonlinear change of variables implied by equations~\eqref{eq:compaction} and \eqref{eq:compaction_second}, including the corresponding Jacobian~\cite{Young:2019yug,DeLuca:2019qsy,Kawasaki:2019mbl,Ferrante:2022hwi}.

Near the collapse threshold, the PBH mass obeys the critical-scaling relation~\cite{Evans:1994pj,Koike:1995jm,Niemeyer:1997mt,Musco:2012au}
\begin{equation}
  m=\kappa M_H\left[\mathcal{C}(r_m)-\Cc\right]^\gamma.
  \label{eq:critical_collapse}
\end{equation}
We adopt the representative radiation-fluid calibration
\begin{equation}
  \kappa=3.3,
  \qquad
  \gamma=0.36.
\end{equation}
The critical exponent $\gamma$ is fixed by the radiation-fluid critical solution, whereas the normalization $\kappa$ depends mildly on the perturbation profile and on the precise convention used to define the threshold and PBH mass~\cite{Niemeyer:1997mt,Musco:2012au}. Here $M_H$ is the horizon mass when the perturbation re-enters the horizon. During radiation domination,
\begin{equation}
  M_H(k)\simeq1.4\times10^{13}\,M_\odot
  \left(\frac{g_*}{106.75}\right)^{-1/6}
  \left(\frac{k}{\mathrm{Mpc}^{-1}}\right)^{-2}.
  \label{eq:horizon_mass}
\end{equation}

The mass fraction collapsing into PBHs is evaluated by integrating over maxima above the threshold and with negative curvature,
\begin{equation}
  \beta(m)=\int_{\Cc}^{\infty}\dif\mathcal{C}(r_m)
  \int_{-\infty}^{0}\dif\mathcal{C}''(r_m)\,
  \frac{m}{M_H}
  P\!\left[\mathcal{C}(r_m),\mathcal{C}''(r_m)\right],
  \label{eq:beta}
\end{equation}
where the PBH mass appearing in the integrand is determined by \Eq{eq:critical_collapse}. This form implements threshold statistics for compaction function maxima and includes the broadening of the PBH mass distribution generated by critical collapse~\cite{Young:2019yug,Yoo:2020dkz,Ferrante:2022hwi}.

For PBHs that survive to the present epoch, the differential dark matter fraction is
\begin{equation}
  \fpbh(m)=\frac{1}{\Omega_{\mathrm{CDM}}}
  \left(\frac{M_{\mathrm{eq}}}{m}\right)^{1/2}\beta(m),
  \qquad
  M_{\mathrm{eq}}\simeq2.8\times10^{17}\,M_\odot,
  \label{eq:fpbh}
\end{equation}
Here $\Omega_{\mathrm{CDM}}$ denotes the present-day cold dark matter (CDM) density parameter.
The mass function is normalized according to
\begin{equation}
  \int \fpbh(m)\,\dif\ln m=\fpbh.
  \label{eq:fpbh_norm}
\end{equation}
The conversion from the formation fraction to the present-day mass function follows from the different redshift scalings of the PBH and radiation energy densities between formation and matter--radiation equality~\cite{Young:2019yug,Yoo:2020dkz}. In summary, equations~\eqref{eq:joint_pdf}--\eqref{eq:fpbh} provide the mapping from the posterior on $(A,k_*,\alpha,\beta)$ to the PBH abundance.

\section{Signal-recovery simulation}
\label{sec:simulation}
To assess the capability of Taiji to constrain the curvature spectrum described above,
we perform an end-to-end Bayesian analysis using simulated data generated by \texttt{Triangle}~\cite{Du:2025xdq}. The simulator constructs six inter-spacecraft $\eta$ observables with realistic orbital dynamics. For noise simulation, we include test-mass acceleration (ACC) noise and optical metrology system (OMS) noise, whose power spectral densities are
\begin{equation}
  \begin{aligned}
    S_{\mathrm{acc},ij}(f)&=A_{\mathrm{acc},ij}^2
      \left[1+\left(\frac{0.4\,\mathrm{mHz}}{f}\right)^2\right]
      \left[1+\left(\frac{f}{8\,\mathrm{mHz}}\right)^4\right]
      \left(\frac{1}{2\pi fc}\right)^2,\\
    S_{\mathrm{oms},ij}(f)&=A_{\mathrm{oms},ij}^2
      \left[1+\left(\frac{2\,\mathrm{mHz}}{f}\right)^4\right]
      \left(\frac{2\pi f}{c}\right)^2.
  \end{aligned}
  \label{eq:noise_psd}
\end{equation}
Here $ij$ labels a movable optical sub-assembly. The nominal amplitudes are $A_{\mathrm{acc}}=3\,\mathrm{fm}\,\mathrm{s}^{-2}/\sqrt{\mathrm{Hz}}$ and $A_{\mathrm{oms}}=8\,\mathrm{pm}/\sqrt{\mathrm{Hz}}$, and the link-dependent amplitudes are allowed to fluctuate by $20\%$ around these values.

Laser-frequency noise is suppressed by second-generation time-delay interferometry (TDI). A generic TDI observable is a linear combination of delayed $\eta$ measurements,
\begin{equation}
  \Delta=\sum_{ij}\mathcal{P}_{ij}\eta_{ij},
\end{equation}
where $\mathcal{P}_{ij}$ is a polynomial of delay operators. The strain power spectral density of an isotropic stochastic gravitational-wave background (SGWB) is related to its energy density by
\begin{equation}
  S_h(f)=\frac{3H_0^2}{2\pi^2f^3}\Omega_{\GW}(f).
  \label{eq:Sh}
\end{equation}
The simulated isotropic signal is generated from sources uniformly distributed over the sky.

In addition to instrumental noise and SIGWs, unresolved Galactic binaries form a persistent confusion foreground. The Taiji Data Challenge II catalog identifies resolvable binaries through an iterative subtraction procedure with a signal-to-noise ratio threshold of $7$~\cite{Liu:2023qap}; we use the residual foreground supplied with the prepared dataset. For a pair of TDI channels, the cross-spectral density satisfies
\begin{equation}
  \left\langle s_I(f)s_J^*(f')\right\rangle
  =\frac12\delta(f-f')\left[N_{IJ}(f)+\Gamma_{IJ}(f)S_h(f)\right],
  \label{eq:tdi_csd}
\end{equation}
where $N_{IJ}$ is the noise cross-spectrum and $\Gamma_{IJ}$ is the overlap reduction function. We use the $A$, $E$, and $T$ TDI channels.

The time dependence of the constellation and the anisotropy of the Galactic foreground break the exact stationarity assumed in \Eq{eq:tdi_csd}. A coordinated treatment would divide the observation into weekly quasi-stationary segments~\cite{Jiang:2026lik}, but this is computationally costly. For this forecast, we follow the simplified treatment of~\cite{Wu2025}. After segmentation and Fourier transformation, we use the averaged periodogram
\begin{equation}
  S(f)=\frac1n\sum_{i=1}^{n}s_i^\dagger(f)s_i(f),\quad s_i(f) = \left[s_{i,A}(f)\;s_{i,E}(f)\;s_{i,T}(f)\right],
  \label{eq:periodogram}
\end{equation}
where $i$ labels the segments. The likelihood is approximated by
\begin{equation}
  \ln p(s|\bar{C})=-n\sum_f\left\{
  \ln|\bar{C}(f)|+\frac{2}{\tau}\mathrm{Tr}\left[S(f)\bar{C}^{-1}(f)\right]
  \right\},
  \label{eq:likelihood}
\end{equation}
up to an additive constant, where $\bar{C}(f)$ is the averaged model covariance matrix and $\tau$ is the duration of a segment. Since $\langle S(f)\rangle=\bar{C}(f)$, the approximation is unbiased in its first moment.

We perform the Bayesian analysis using a Markov chain Monte Carlo (MCMC) sampler.
The sampler employs the affine-invariant stretch-move proposal~\cite{camcos2010}.
We do not assume a specific spectral model for the confusion foreground; instead,
its cross-channel correlations are estimated directly from the simulated data.
The instrumental-noise amplitudes and the four BPL parameters are inferred jointly. The detector data are analyzed over $10^{-4}$--$0.05\,\mathrm{Hz}$. We first validate the pipeline with an injected spectrum,
\begin{equation}
  A=2\times10^{-3},\quad
  f_*=4\times10^{-3}\,\mathrm{Hz},\quad
  \alpha=3,\quad
  \beta=5,
  \label{eq:injection}
\end{equation}
whose peak lies inside the sensitive band. We then analyze a pure-noise realization to obtain projected upper limits. To map the two edges of the curvature-scale reach into PBH masses, we perform separate peak-frequency scans over $0.04$--$0.4\,\mathrm{mHz}$ and $0.1$--$4\,\mathrm{Hz}$. In these scans the peak can lie outside the detector band; Taiji then constrains the appropriate power-law tail of the SIGW spectrum.

\section{Results and discussion}
\label{sec:results}

\subsection{Constraints on the BPL spectrum}
\begin{figure}[H]
  \centering
  \includegraphics[width=.9\columnwidth]{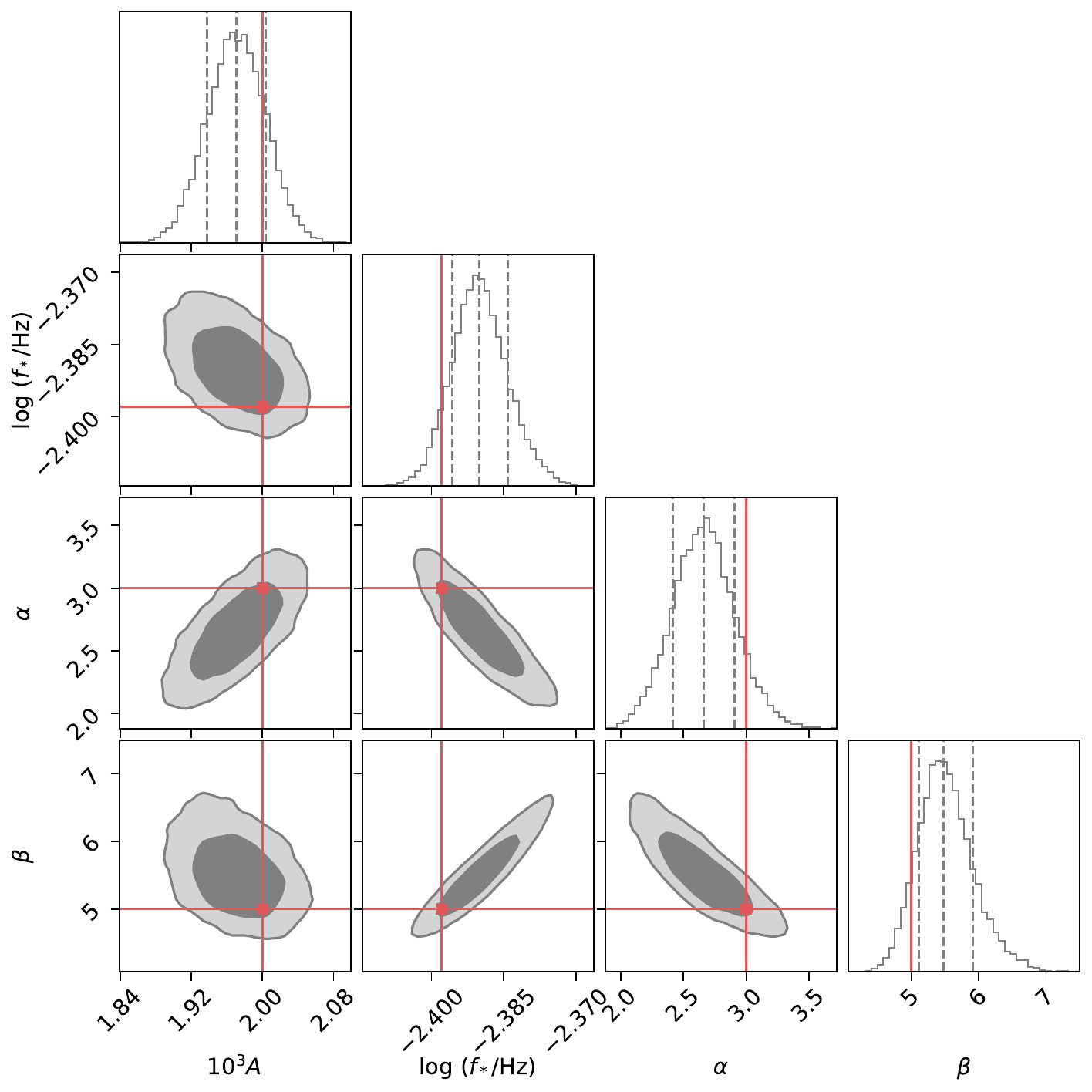}
  \caption{Posterior distributions of the BPL parameters in \Eq{eq:bpl} for the injected signal. The shaded contours denote the $68\%$ and $95\%$ credible regions. The red lines mark the injected values given in \Eq{eq:injection}.}
  \label{fig:posterior_inj}
\end{figure}

We first examine the recovery performance for the injected signal. The resulting posterior distribution is shown in figure~\ref{fig:posterior_inj}, where the injected values lie within the $68\%$ credible contours. The approximate likelihood given by \Eq{eq:likelihood} appears adequate for this recovery test.

\begin{figure}[H]
  \centering
  \includegraphics[width=.9\columnwidth]{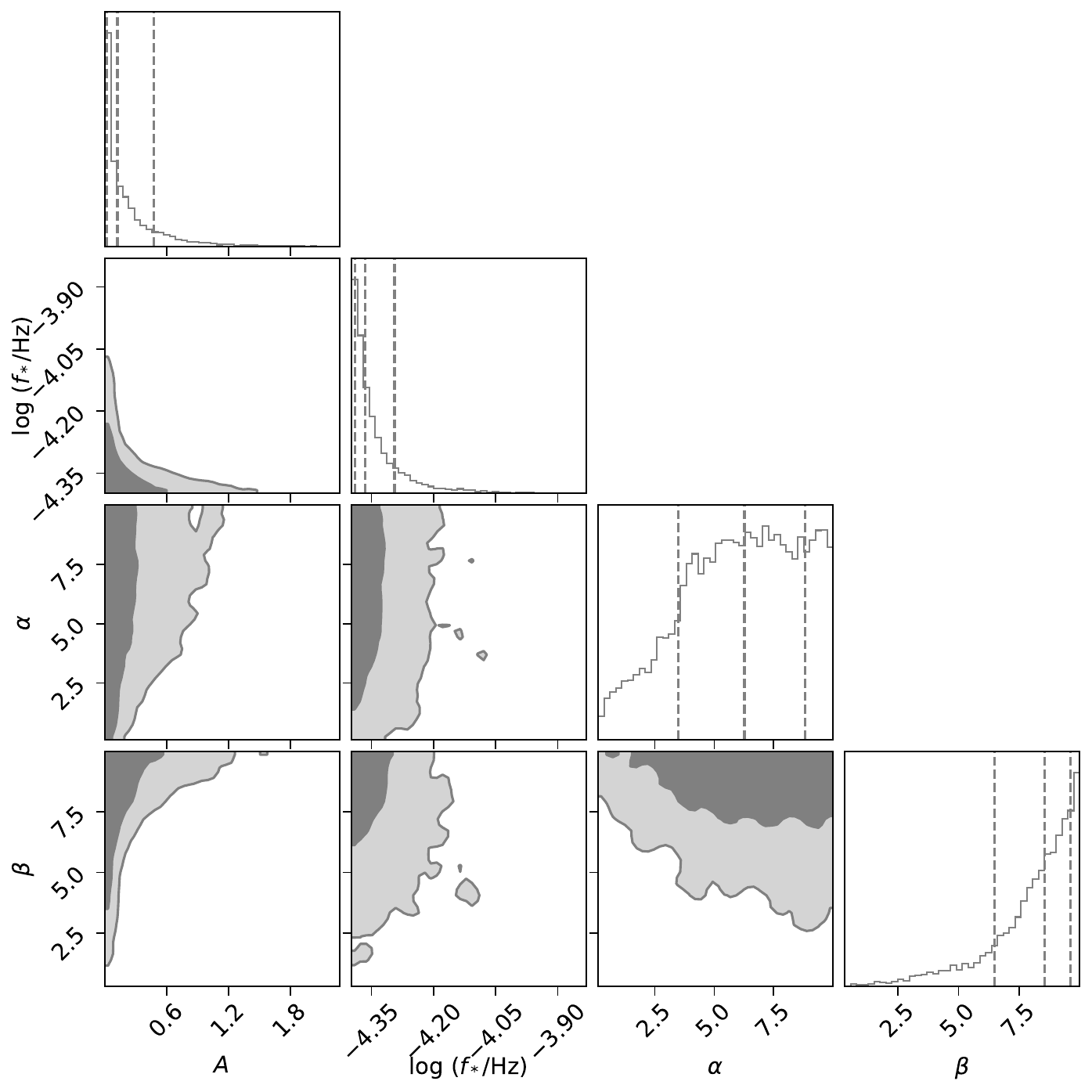}
  \includegraphics[width=.9\columnwidth]{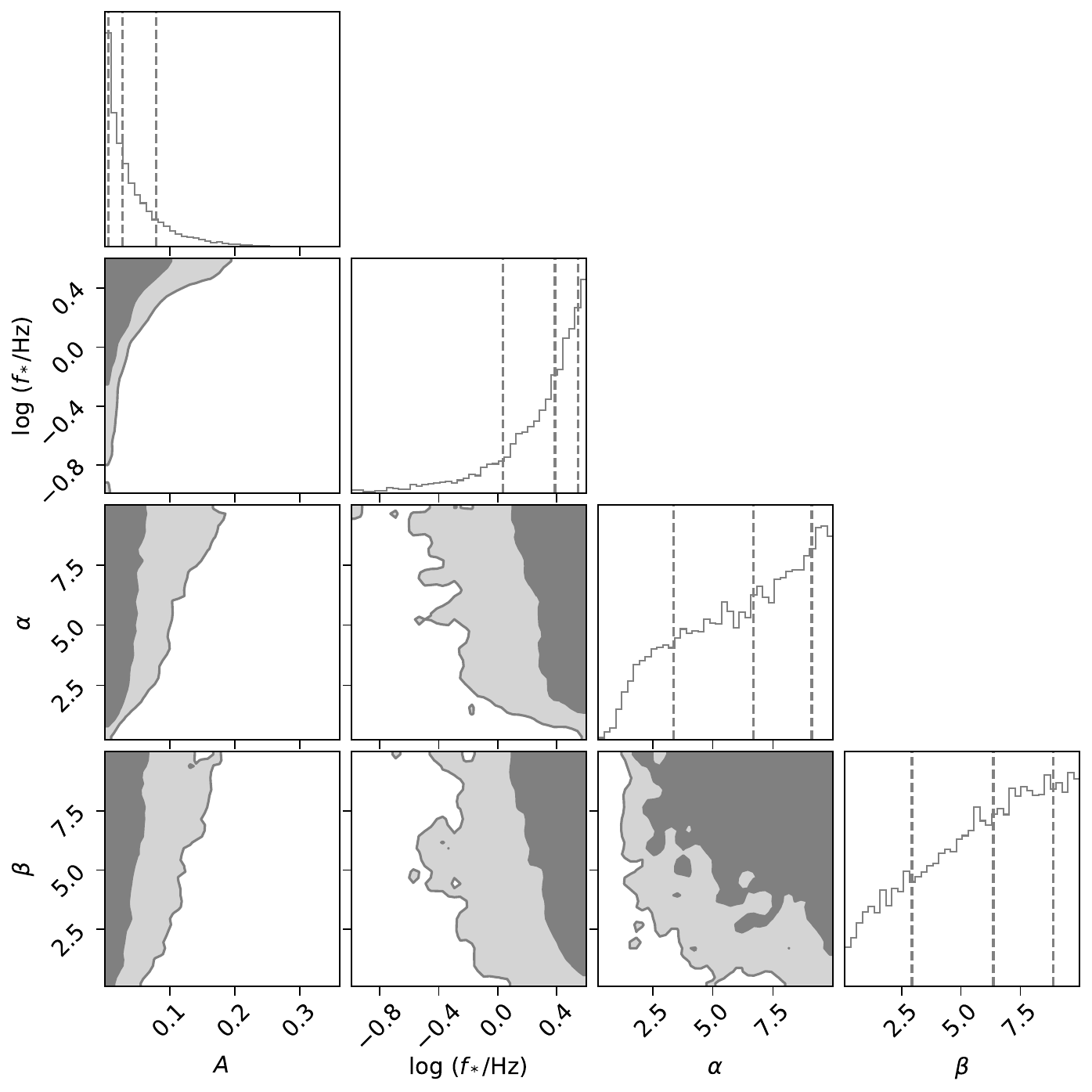}
  \caption{Posterior distributions of the BPL parameters in \Eq{eq:bpl} for the low-frequency scan (top) and high-frequency scan (bottom), obtained from a pure-noise realization. The shaded contours denote the $68\%$ and $95\%$ credible regions. The weak constraints on $\alpha$ and $\beta$ reflect their degeneracy with the amplitude when only a spectral tail is sampled in the Taiji band.}
  \label{fig:posterior}
\end{figure}

Figure~\ref{fig:posterior} shows the posterior distributions obtained from the pure-noise realization for the low- and high-$f_*$ scans. No localized posterior mode indicative of an SGWB signal is present. The slope parameters remain weakly constrained because the data primarily limit the combination of the amplitude and the tail that enters the Taiji band. The degeneracy is especially pronounced when the spectral peak is far outside the analyzed frequency range.

The ordering of the amplitude limits follows the tail sampled by the detector. For the low-frequency scan, the BPL peak lies below the Taiji band and the observed frequencies probe the ultraviolet tail controlled mainly by $\beta$; the marginalized $95\%$ upper limit on $A$ is typically of order unity. For the high-frequency scan, the detector probes the infrared tail controlled mainly by $\alpha$, and the corresponding upper limit is of order $0.1$ \cite{Yuan:2019wwo,Li:2024lxx}. These limits should not be interpreted as pointwise bounds on $\Pzeta(k)$ independent of the assumed BPL shape, because changes in $(\alpha,\beta)$ can compensate changes in $A$.

\subsection{Projected constraints on the PBH abundance}
\begin{figure}[H]
  \centering
  \includegraphics[width=.9\columnwidth]{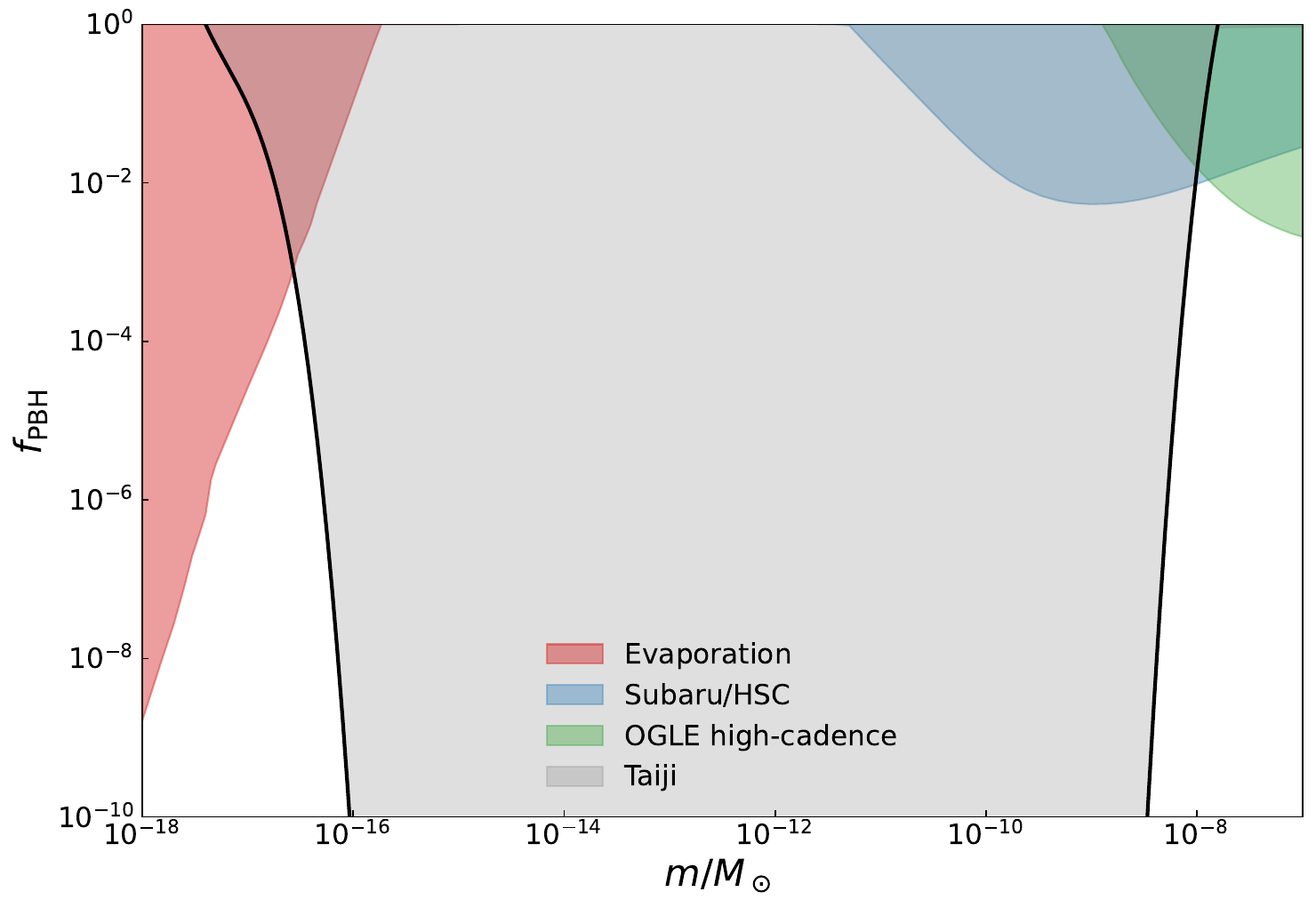}
  \caption{$95\%$ upper limits on the PBH fraction inferred from the Taiji null-SIGW analysis. The grey shaded region is the Taiji exclusion. For comparison, we also show representative existing constraints from the combined Hawking-evaporation envelope, Subaru/HSC microlensing, and the OGLE high-cadence microlensing limit.}
  \label{fig:pbh}
\end{figure}

We propagate each posterior sample through the PBH formation calculation in section~\ref{subsec:pbh}, obtaining a corresponding pair $(M_{\rm PBH},\fpbh)$. We then sort the posterior samples by $f_*$ and divide them into $80$ groups with approximately equal numbers of samples. For each group, the representative PBH mass is taken as the median of $M_{\rm PBH}$, while the corresponding upper limit on $\fpbh$ is taken as the $95$th percentile of the $\fpbh$ samples within the group. Each group contains more than $400$ samples, which suppresses large statistical fluctuations.
The resulting scan points are summarized in figure~\ref{fig:pbh}. Near Taiji's most sensitive frequencies, the inferred PBH upper limits fall below the machine limit and return zero. We therefore retain explicit scan points only near the two edges. The resulting $95$th-percentile upper limits satisfy $\fpbh^{95\%}<1$ for PBH masses from approximately $6.2\times10^{-18}\,M_\odot$ to $1.7\times10^{-8}\,M_\odot$, excluding PBHs as all of the dark matter throughout this interval under the adopted formation prescription. Upper limits above unity do not constrain a physical PBH dark matter fraction and we only show the results $\fpbh\le 1$.

For the high-$f_*$ scan, the lower-mass edge of the region satisfying $\fpbh^{95\%}<1$ occurs at approximately $6.2\times10^{-18}\,M_\odot$, where the upper limit approaches unity. This lower-mass region overlaps with the Hawking evaporation constraints. Under standard Hawking evaporation, PBHs well below $10^{-18}\,M_\odot$ do not survive to the present day. However, the memory burden effect can suppress the late-time evaporation and allow ultralight PBHs to survive~\cite{Jiang_2024,Dvali:2024hsb}.

For the low-$f_*$ scan, $f_*=0.04$--$0.4\,\mathrm{mHz}$, the upper-mass edge of the region satisfying $\fpbh^{95\%}<1$ occurs at approximately $1.7\times10^{-8}\,M_\odot$, where the upper limit approaches unity. At the lower-mass side of this branch, the projected Taiji bound falls below the current Subaru Hyper Suprime-Cam (HSC) and recent Optical Gravitational Lensing Experiment (OGLE) high-cadence microlensing bounds. The forecast remains competitive up to masses of order $10^{-8}\,M_\odot$, but then weakens rapidly.

The external bounds in figure~\ref{fig:pbh} assume monochromatic PBH populations, whereas the BPL spectra considered here generate extended mass functions through critical collapse. The comparison between the Taiji forecast and other current constraints should therefore be interpreted as an order-of-magnitude indication.

\section{Conclusion}
We have forecast the capability of Taiji to constrain enhanced primordial curvature perturbations and the associated PBH abundance through a search for SIGWs. The curvature spectrum is modeled by a BPL model, and the stochastic signal is analyzed jointly with instrumental noise and a residual Galactic-binary foreground. A signal injection validates the pipeline, while a pure-noise realization yields the projected null-signal limits discussed here.

To translate the curvature spectrum posterior into a PBH abundance, we use the nonlinear compaction function, a shape-dependent collapse threshold, critical collapse, and the two-dimensional joint Gaussian distribution of the compaction function. This treatment takes into account the nonlinear effects during the PBH formation.

The forecast yields $\fpbh^{95\%}<1$ for PBHs with masses from approximately $6.2\times10^{-18}\,M_\odot$ to $1.7\times10^{-8}\,M_\odot$, thereby excluding PBHs as all of the dark matter across this interval under the adopted formation prescription. The null detection of SIGWs leads to stringent constraints on the PBH abundance, $\fpbh^{95\%}\le10^{-10}$ near Taiji's most sensitive frequency band. On the asteroid-mass side, the projected constraint improves on both the Subaru/HSC and recent OGLE high-cadence microlensing bounds over part of the relevant range. In the low-mass range, the constraints extend to ultralight PBHs range where these PBHs have already evaporated through standard Hawking radiation. However, if taking into account the mechanism such as memory burden effect that would suppress PBH evaporation, then our results can be interpreted as another independent method to search or constrain the PBH dark matter.

Future work could include more realistic noise modeling and a more complete treatment of the anisotropic Galactic foreground. Primordial non-Gaussianity can also change the relation between SIGWs and PBH abundance~\cite{Yuan:2020iwf,Meng:2022ixx,Yuan:2023ofl,Liu:2023ymk}. With these improvements, Taiji will provide a complementary probe of primordial structure on scales far smaller than those accessible to conventional cosmological observations.

The MCMC sampler used in this work is available at \url{https://github.com/DrizzleatDusk/TaijiSampler}.

\section*{Acknowledgments}
Y.J. is supported by the China Postdoctoral Science Foundation under Grant Number 2025M783376. Q.-G.H. is supported by the National Natural Science Foundation of China (NSFC) grants No.~12547110, 12475065, and 12447101.

\bibliographystyle{scpma}
\bibliography{refs.bib}

\end{multicols}
\end{document}